\documentclass[
    ,final            
  ]
  {aipproc}

\usepackage{psfig}
\usepackage{graphicx}

\layoutstyle{6x9}

\begin{document}

\title{Jet Studies in STAR via 2+1 Correlations}


\author{Hua Pei, for the STAR collaboration}{
  address={Post-doc research associate, 845 W Taylor St, MC 273, Dept. of Physics, University of Illinois at Chicago, Chicago, IL, 60607, U.S.A.}
}

\begin{abstract}

This paper reports analysis on jet-medium interactions and di-jet surface emission bias at RHIC, based on a new multi-hadron correlation technique called \texttt{2+1} where back-to-back high $p_{T}$ hadron triggers work as proxy of di-jets. In contrast with traditional correlations with single triggers, the associated hadron distributions and spectra at both trigger sides show no evident modification from d+Au to central Au+Au collisions. This observation stands for both cases when triggers pairs are symmetric($p_T>$5GeV/$c$ vs. $p_T>$4GeV/$c$) or asymmetric($E_T>$10GeV/$c$ vs. $p_T>$4GeV/$c$). 

\end{abstract}

\maketitle


\section{Introduction}

Di-hadron correlation measurements in heavy ion collisions using a single high-$p_{T}$ trigger have observed broader distributions
of associated hadrons on the away-side azimuthal ($\Delta\phi\sim\pi$) \cite{star_corr,phenix_corr} in heavy ion collisions relative to $p+p$ and $d+Au$, and a long-range $\Delta\eta$ plateau called \texttt{ridge} at near-side ($\Delta\phi\sim 0$) \cite{ridge}.
However, they don't identify the away side jet which may endure more jet-medium interaction.
On the other hand, two high-$p_{T}$ hadron correlations exhibit jet-like peaks in both near-/away-sides \cite{o31,phenix_nomodify} with little shape modification from d+Au to central Au+Au, but a strong suppression on the away-side amplitude. 
The data may be interpreted in two scenarios: in-medium energy loss followed by in-vacuum fragmentation, and finite probability to escape the medium without interactions.
The analysis in this paper uses a new 3-particle ($2+1$) correlation technique, which measures the correlation of low-$p_{T}$ particle with a pair of back-to-back high-$p_{T}$ trigger pair as proxy of di-jets. The asymmetry between the two triggers $p_{T}$ are varied as an attempt to control the path length each parton travels in the medium, thus the difference between their final energy.

\section{Analysis and Result}


This paper is based on data collected by the STAR (Solenoidal Tracker at RHIC) experiment \cite{STAR} in the years 2003-2004 and 2007-2008 of collisions events center-of-mass per nucleon pair energy of 200 GeV, of both d+Au and Au+Au.
The Au+Au data are divided into multiple centrality bins based on the charged track multiplicity at mid-rapidity ($|\eta|<0.5$). The number of participating nucleons ($N_{part}$) and the number of binary collisions ($N_{coll}$) for each centrality bin used in the analysis are calculated via Monte-Carlo Glauber model.
The high-$p_T$ trigger pairs are required to be back-to-back $|\phi_{trig1}-\phi_{trig2}-\pi|<0.2$ to work as proxy of dijets.
%
The associated hadrons 1.5~GeV/$c$~$<p_T^{assoc}<p_T^{trig1}$) is selected to coincide with the range where the broad away-side and the near-side ridge are reported.
The correlation functions are defined as

$$\frac{d^2N}{d\Delta\eta d\Delta\phi}=\frac{1}{N_{trig} \rm{\epsilon}_{pair}} \left(\frac{d^2N_{raw}}{d\Delta\eta d\Delta\phi}-a_{zyam} \frac{d^2N_{Bg}}{d\Delta\eta d\Delta\phi}\right)
$$
\noindent

$N_{trig}$ is the number of trigger pairs, and ${d^2N_{raw}}/{d\Delta\eta\,d\Delta\phi}$ is the associated hadron distributions
corrected by $\rm{\epsilon}_{pair}$ for single-track efficiency and pair acceptance effects. 
The background $\frac{d^2N_{Bg}}{d\Delta\eta d\Delta\phi}$ is estimated from mixing-event technique after being modulated by the flow \cite{flow}.
The background due to randomly associated triggers as di-jets \cite{3pPRL} is also considered from trigger-trigger correlation in a similar way.
The overall background level $a_{zyam}$ is then decided with the Zero-Yield at Minimum (ZYAM) method \cite{zyam1}. 
The $p_{T}$ spectra for associated hadrons is measured within 0.5~radians in $\Delta\phi$ and 0.5 in $\Delta\eta$ of the respective trigger direction, with background removed.
The systematic uncertainties from all sources and items above are evaluated.
Their sums are highest in central Au+Au events (<20\%) and much lower in d+Au and peripheral Au+Au,
%
%
and strongly correlate between same-/away-sides and mostly cancel out for such comparisons.



\begin{figure*}[hbt]
\centerline{\psfig{file=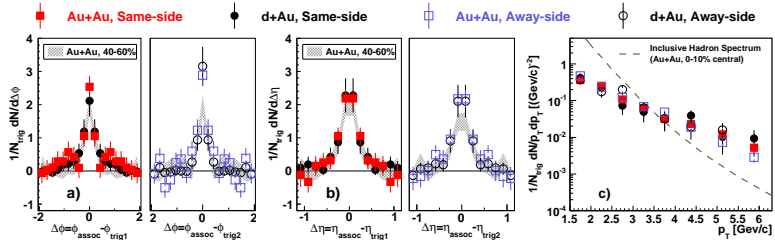,width=0.7\textwidth}}
\vspace*{-0.2cm}
\caption{
Projections of 2+1 correlation on  $\Delta\phi$ ($a$) and $\Delta\eta$ ($b$) 
 for 200 GeV Au+Au  and d+Au data \cite{2plus1_paper}: $5<p_T^{trig1}<10$~GeV/$c$, 4~GeV/$c$~$<p_T^{trig2}<p_T^{trig1}$, $1.5<p_T^{assoc}<p_T^{trig1}$. Errors shown are statistical. 
$c)$ $p_{T}$ spectra per trigger pair for the same-/away-side associated hadrons ($|\Delta\phi|<0.5$, $|\Delta\eta|<0.5$). Errors are the quadrature sum of the statistical and systematic uncertainties.
Inclusive charged hadron spectra from 10\% most central Au+Au data \cite{star_raa} is shown for comparison. 
}
\label{figure1_2plus1_paper}
\end{figure*}

The symmetric trigger work shown in Fig. \ref{figure1_2plus1_paper} is from most recent STAR 2+1 correlation publication \cite{2plus1_paper}. 
The $\Delta\phi$ and $\Delta\eta$ projections (symmetrized about 0) of the correlation function are plotted
in Fig. \ref{figure1_2plus1_paper}$a$ and $b$ respectively.
This measurement constitutes the first observation of not only the near-side, but also the away-side correlation structure in central Au+Au reproducing those of d+Au as a reference (without hot quark matter), in both $\Delta\phi$ and $\Delta\eta$, for the associated hadrons in this kinematic range. 
No evidence of \texttt{dip} or \texttt{ridge} is present. However, statistical limitations prevents a complete exclusion of ridge \cite{ridge}. 
The similarity between Au+Au and d+Au is further supported by their associated hadron $p_{T}$ spectra plotted in Fig. \ref{figure1_2plus1_paper}$c$,
%
on contrary to previous di-hadron correlations measurments in a similar kinematic range, where significant softening of the away-side spectra is observed \cite{star_corr} indicating energy deposition in the medium.
The jet energy is then estimated by summing the $p_{T}$ of trigger and charged associates spectra. 
%
The result in the 12\% central Au+Au data 
$\Delta(Au+Au) = \Sigma(p_T)^{same}-\Sigma(p_T)^{away}=1.59\pm 0.19$~GeV/$c$, similar to the minimum bias d+Au data value of $\Delta(d+Au) = 1.65 \pm 0.39$~GeV/$c$. 
This number is close to the initial state kinematic effects 1.6~GeV/$c$ \cite{renk_kt} and disfavors additional partons losing energy into medium in Au+Au case.

\begin{figure*}[thb]
\includegraphics[width=0.7\textwidth]{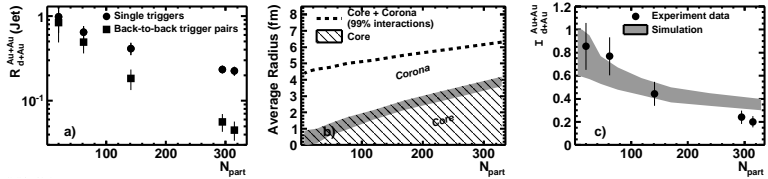}
\caption{\label{figure2_2plus1_paper}Comparison between data and model \cite{2plus1_paper}. $a)$ Relative production rates for jets and di-jets in Au+Au of different centralities with respect to d+Au. $b)$ Calculations of the Glauber-based core/corona  model that accommodates inclusive hadron suppression and single jet production rate results. $c)$ Conditional di-jet survival probability in Au+Au data compared to d+Au reference. The points are from data. The shadow band shows the expectations from Glauber-based core/corona model described in the text. Error bars in $a)$ and $c)$ are the quadrature sum of the statistical and systematic uncertainties.}
\end{figure*}

As a conclusion, the $2+1$ results are consistent with lack of medium-induced effects on those di-jets selected by this analysis, and favors a surface jet emission model.
%
This model is then tested by
measuring the nuclear modification factors $R_{d+Au}^{Au+Au}$ (ratio of binary-scaled per-event trigger counts in Au+Au and d+Au data) for the primary (single) triggers and di-jet triggers as shown in Fig. \ref{figure2_2plus1_paper}$a$.

%


%
A Monte Carlo Glauber model based on surface emission is applied, assuming the medium in heavy-ion collisions consists of a completely opaque core (full jet absorbtion) surrounded by a permeable corona (no jet-medium interactions).
The relative size of the core is estimated from the $R_{AA}$ of single triggers \cite{phenix_raa,star_raa} because they are purely from the corona and bypass the core in this model. The calculation is shown in Fig.\ref{figure2_2plus1_paper}$b$. 
%
The model then calculates the conditional survival probability of di-jets $I_{d+Au}^{Au+Au} =\frac{R_{AA} (trigger-pairs)} {R_{AA}(single-triggers)}$ and compares with the experimental data, shown in Fig. \ref{figure2_2plus1_paper}$c$.
The data are symbol points, and the expected rates from core/corona model are shown as a band, where the width reflects the uncertainty in the published $R_{AA}$ data.
They agree with each other qualitatively.
%
However, core emission where neither of the di-jets interacted with the medium cannot be ruled out by this analysis.


As for the path-length dependent energy-loss models where in-medium energy loss followed by in-vacuum fragmentation,
trigger pairs of similar $p_{T}$
can possibly bias towards partons travel similar path length of medium and lose similar energy.
Thus highly asymmetric trigger pairs in $p_{T}$ are selected as proxy of di-jets of big asymmetry in energy loss.
The primary triggers are BEMC tower clusters of $10<E_T^{trig1}<15$~GeV dominated by decayed $\pi^{0}$s.
The direct$-\gamma$s contamination is non negligible \cite{phenix_direct_gamma}
but won't affect the validity of comparison, as direct$-\gamma$s don't lose energy in the medium and away-side parton will lose relatively more energy if path-length dependence stands.
%

\begin{figure}
\centerline{\psfig{file=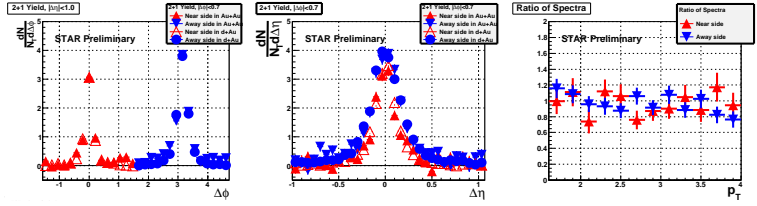,width=0.7\textwidth}}
\vspace{-0.5cm}
\caption{\label{figure3_asymmetric_trigger}2+1 correlation in asymmetric trigger case. Projections of 2+1 correlation for 200 GeV Au+Au and d+Au data on $a)$ $\Delta\phi$ and $b)$ $\Delta\eta$: $10<E_T^{trig1}<10$~GeV, 4~GeV/$c$~$<p_T^{trig2}<10$~GeV/$c$, $1.5<p_T^{assoc}<10$~GeV/$c$. $c)$ Au+Au spectra divided by d+Au. Errors shown are statistical.}
\end{figure}

The correlations are shown in Fig. \ref{figure3_asymmetric_trigger} $a$ and $b$. Similar jet-like peak shapes and magnitudes sustain from d+Au to central Au+Au collisions at both same-/away-sides. Again, no evident ridge or dip structure is observed
in Au+Au data.
The away-side magnitude is higher than near-side in both Au+Au and d+Au, which can be due to either the direct-$\gamma$ contamination in primary trigger, or momentum conservation at lower-$p_{T}$ trigger (away-)side if back-to-back jets are of similar energy. 
On the other hand, such difference varies little from d+Au to Au+Au, indicating no evidence of asymmetry in energy loss of surviving parton pairs expected from path-length dependence model \cite{renk_kt}.
%
%
The ratios of Au+Au associates spectra at either trigger side divided by d+Au are plotted in Fig. \ref{figure3_asymmetric_trigger}$c$.
These ratios are consistent with flat and close to unity at either trigger side, showing no evidence of softening or strong suppression,
and further supports the assumption of strong surface-bias whenever a jet-like structure is observed.

\section{Summary}

The mechanisms of jet-medium interactions was investigated using a novel technique called $2+1$ correlations, studying low-$p_{T}$ hadrons associated with a correlated pair of back-to-back high $p_{T}$ particles as proxy of di-jets.
Both same-/away-side correlations are found similar from d+Au to central Au+Au data, and so is the associated hadron spectra on each trigger side, in the kinematic range selected of $p_T^{assoc}>1.5$~GeV/$c$.
%
%
The path-length dependence energy loss models expect that in either symmetric or very asymmetric trigger pairs cases such similarity shall be broken for di-jets if they are from deep within the medium (non-tangential), which wasn't observed at this stage of analysis within the errors.
Meanwhile, systematic assessment of di-jet production rates supports tangential emission bias in a simplistic core/corona scenario.
%





\end{document}